tex

\input harvmac
\input epsf
\def\journal#1&#2(#3){\unskip, \sl #1\ \bf #2 \rm(19#3) }
\def\andjournal#1&#2(#3){\sl #1~\bf #2 \rm (19#3) }

\def\frac#1#2{{#1\over#2}}

\def\vev#1{\langle#1\rangle}

\def\inbar{\,\vrule height1.5ex width.4pt depth0pt}
\def\IC{\relax\hbox{$\inbar\kern-.3em{\rm C}$}}
\def\IR{\relax{\rm I\kern-.18em R}}
\def\IP{\relax{\rm I\kern-.18em P}}
\def\IZ{\relax{\rm I\kern-.18em * 0<
* 1
* 2
* 3
* 4
* 5
* 6
* 7
* 8
* 9
*10
*11
*12
*13
*14
*15
Z}}

%
%

%
\catcode`\@=11
\def\slash#1{\mathord{\mathpalette\c@ncel{#1}}}
\overfullrule=0pt

\def\underrel#1\over#2{\mathrel{\mathop{\kern\z@#1}\limits_{#2}}}

\catcode`\@=12


%

\def\vev#1{\left\langle #1 \right\rangle}


\def\j{{\bf j}}

\rightline{RI-11-03} \Title{ \rightline{hep-th/0311188}}
{\vbox{\centerline{Probing Orientifold Behavior Near NS Branes}}}
\medskip
\centerline{\it Dmitri Burshtyn, Shmuel Elitzur and Yaakov Mandelbaum}
\bigskip
\smallskip
\centerline{Racah Institute of Physics, The Hebrew University}
\centerline{Jerusalem 91904, Israel}
\smallskip

\bigskip\bigskip\bigskip
\noindent The effect of NS $5$ branes on an orientifold is
studied. The orientifold is allowed to pass through a pile of $k$
NS branes forming a regularized CHS geometry. Its effect on open
strings in its vicinity is used to study the change in the
orientifold charge induced by the NS branes.

\vfill \Date{12/03}
 \lref\givkut{ For a review see A.~Giveon and D.~Kutasov,
Rev.\ Mod.\ Phys.\  {\bf 71}, 983 (1999) [arXiv:hep-th/9802067].
}
\lref\tsabar{ S.~Elitzur, A.~Giveon, D.~Kutasov and D.~Tsabar,
Nucl.\ Phys.\ B {\bf 524}, 251 (1998) [arXiv:hep-th/9801020].
}
\lref\chs{ C.~G.~Callan, J.~A.~Harvey and A.~Strominger,
arXiv:hep-th/9112030.
} \lref\pol{J. Polchinski, String Theory, Cambridge University
Press 1998, vol.I, p. 191}
\lref\ah{For a review see O.~Aharony,
Class.\ Quant.\ Grav.\  {\bf 17}, 929 (2000)
[arXiv:hep-th/9911147].
}
\lref\ogvaf{ H.~Ooguri and C.~Vafa,
Nucl.\ Phys.\ B {\bf 463}, 55 (1996) [arXiv:hep-th/9511164].
}
\lref\sfet { C.~Csaki, J.~Russo, K.~Sfetsos and J.~Terning,
``Supergravity models for 3+1 dimensional {QCD},'' Phys.\ Rev.\ D
{\bf 60}, 044001 (1999) [arXiv:hep-th/9902067].
}
\lref\gkp{ A.~Giveon, D.~Kutasov and O.~Pelc,
JHEP {\bf 9910}, 035 (1999) [arXiv:hep-th/9907178].
}
\lref\gk{ A.~Giveon and D.~Kutasov,
JHEP {\bf 0001}, 023 (2000) [arXiv:hep-th/9911039].
}
\lref\huss{ L.~R.~Huiszoon, K.~Schalm and A.~N.~Schellekens,
Nucl.\ Phys.\ B {\bf 624}, 219 (2002) [arXiv:hep-th/0110267].
}

\lref\prad{ G.~Pradisi, A.~Sagnotti and Y.~S.~Stanev,
Phys.\ Lett.\ B {\bf 354}, 279 (1995) [arXiv:hep-th/9503207].
}
\lref\evans{ N.~Evans, C.~V.~Johnson and A.~D.~Shapere,
Nucl.\ Phys.\ B {\bf 505}, 251 (1997) [arXiv:hep-th/9703210].
}
\lref\clif{ C.~V.~Johnson,
Phys.\ Rev.\ D {\bf 56}, 5160 (1997) [arXiv:hep-th/9705148].
}
\lref\bach{ C.~Bachas, N.~Couchoud and P.~Windey,
JHEP {\bf 0112}, 003 (2001) [arXiv:hep-th/0111002].
}

\lref\brunner{ I.~Brunner,
JHEP {\bf 0201}, 007 (2002) [arXiv:hep-th/0110219].
}
\lref\couch{ N.~Couchoud,
JHEP {\bf 0203}, 026 (2002) [arXiv:hep-th/0201089].
}
\lref\forge{ S.~Elitzur, A.~Forge and E.~Rabinovici,
Nucl.\ Phys.\ B {\bf 359}, 581 (1991).
}
\lref\mandal{ G.~Mandal, A.~M.~Sengupta and S.~R.~Wadia,
Mod.\ Phys.\ Lett.\ A {\bf 6}, 1685 (1991).
}
\lref\wit{ E.~Witten,
Phys.\ Rev.\ D {\bf 44}, 314 (1991).
}
\lref\sag{ A.~Sagnotti and Y.~S.~Stanev,
Fortsch.\ Phys.\  {\bf 44}, 585 (1996) [Nucl.\ Phys.\ Proc.\
Suppl.\  {\bf 55B}, 200 (1997)] [arXiv:hep-th/9605042].
}
\lref\forste{ S.~Forste, D.~Ghoshal and S.~Panda,
Phys.\ Lett.\ B {\bf 411}, 46 (1997) [arXiv:hep-th/9706057].
}

\newsec{Introduction}

NS $5$ branes are non perturbative string configurations.
Therefore it is not easy to study the behavior of fundamental
strings in their vicinity. Fundamental strings moving far enough
from the core of the NS brane are influenced just by the NS $3$
flux emanating from it. The behavior of such strings on top of the
core is a more difficult problem.

One of the interesting effects of a NS brane is the change of sign
of the RR charge of a $6$ orientifold passing through it. As was
first noted by Evans, Johnson and Shapere \evans,  when a NS $5$
brane divides an $O6$ orientifold into two parts, these parts have
opposite signs \givkut . This effect is required for consistency
of low energy limit field theories built on such a configuration.
It is also easy to see a change of the sign between crosscup
diagrams on both sides of the NS brane, far enough from it, as a
result of the NS $3$ flux of it \tsabar . Here also the
perturbative treatment relies on the fundamental string diagrams
put far from the NS brane.

 String theory in the near horizon of a  stack of $k
NS$ branes is known as Little String Theory \ah. In general it is
not perturbatively  controllable. For $k>1$, the  geometry formed
around the branes \chs\ has in fact the form of  an exact CFT
which is the product $R^{1,5}\times SU(2)_k \times R_\phi$. Here
$R^{1,5}$ is the world volume of the stack of branes, $SU(2)_k$ is
the $S^3$ sphere of angular coordinates around it and $R_\phi$ is
the radial coordinate $\sim e^{\phi}$ away from it. However, there
is a linear dependence of the dilaton field along $\phi$. The non
perturbative nature of the configuration is reflected by the
blowing up of the string coupling at the center, $\phi \to
-\infty$, caused by the linear dilaton.

Use can still be made of the exact CFT form of this background if
this divergence of the string coupling is properly regularized
\ogvaf,\sfet,\gkp,\gk. This can be done by turning on a super
Liouville potential which shields the strong coupling region.
Alternatively, the super Liouville system can be replaced by the
cigar shaped geometry \forge,\mandal,\wit\ of the coset
$SL(2,R)/U(1)$, where the strong coupling region is cut off
geometrically. This type of
\lref\gor{ S.~Elitzur, A.~Giveon, D.~Kutasov, E.~Rabinovici and
G.~Sarkissian,
JHEP {\bf 0008}, 046 (2000) [arXiv:hep-th/0005052].
} regularization corresponds to letting the NS branes being
distributed along some circle rather than sitting on top of each
other. Such a background allows for a perturbative treatment of
both closed \gk\ and open \gor\ strings in the near horizon
neighborhood.

In this paper we use this regularized description to probe the
influence of NS branes on an orientifold. An $O6$ will be put into
the CFT background described above \clif,\forste. The world volume
of this orientifold intersects the $SU(2)$ manifold at two points.
A probe of N $D4$ branes is further connected to the pile of $k$
NS branes. The strings connecting these $D$ branes to their $O6$
images will be studied. As this probe of $D4$ branes is rotated
with respect to the orientifold, passing the equator of the $S^3$
sphere, a phase transition is encountered in the gauge theory on
these $D$ branes. For odd $k$ it will be a transition from an
$SO(N)$ to an $Sp({N\over 2})$ gauge group (or vice versa,
depending on the original sign of the orientifold). For even $k$
the gauge group stays the same on both sides of the transition.
This is consistent with the expectation that each NS brane causes
a change of the sign of the orientifold. The presence of a
$(-1)^k$ factor between the charges of the two intersections of
the orientifold with the $SU(2)$ manifold has been deduced before
in \bach\ from flux arguments. Here we focus on studying the
behavior of open strings and D branes systems under this
transition.
\lref\lll{ K.~Landsteiner, E.~Lopez and D.~A.~Lowe,
JHEP {\bf 9802}, 007 (1998) [arXiv:hep-th/9801002].
} In \lll\ and \tsabar\ such a configuration was studied for the
case $k=1$. The passing from $SO(N)$ to $Sp({N\over 2})$ and the
enhanced gauge symmetry at the transition point were guessed there
on the basis of the known change of sign of the orientifold
induced by the NS brane. Here, for $k>1$, the regularized
background enables one to follow this process not only far from
the NS branes but throughout the transition region.

In sec. $2$ an orientifold is put into a WZW $SU(2)_k$ model. This
system was analyzed algebraically in \prad\ and \sag, and more
geometrically in \brunner, \bach\ and \huss. See also \couch\ for
a discussion of the $SO(3)$ case. The action of the orientifold
$Z_2$ gauging on the symmetry generators in the space of open
strings connecting a pointlike $D$ brane to its mirror image is
studied. In sec. $3$ the treatment in \gor\ of open strings in the
regularized background of a stack of $k$ NS branes is reviewed. In
sec. $4$ an orientifold is put into this background. The results
of the previous sections are used to identify a phase transition
as the $D4$ branes are moved along the configuration. This
transition is connected to the change of the sign of the
orientifold induced by the $NS$ branes. Sec. $5$ is a conclusion.

\newsec{Orientifold Action on Symmetry Generators}

 The action of an $SU(2)_k$ WZW model on a world sheet $\Sigma$ is
\eqn\wzw{S={ k\over{4 \pi}}[\int_{\Sigma}d^2z L^{kin} + \int_B
\omega^{WZ}]}
where  $L^{kin} = Tr  (\partial_z g \partial_{\bar z
} g^{-1} )$ ,
 $\omega^{WZ}={ 1\over 3} Tr (g^{-1} dg)^3$, and $B$ is a $3$ manifold
 bounded by $\Sigma$.  $g(z,\bar z)$ is the embedding of
 $\Sigma$ into  $SU(2)$  group manifold. The variation

\eqn\var{\delta g= \epsilon_L g - g \epsilon_R} $\epsilon_{L,R}$
being $z,\bar z$ dependent infinitesimal group generators, induces
the following variation in the action:

\eqn\dels{\delta S={k\over{2\pi}}\int_{\Sigma}d^2 z Tr[\epsilon_L
\bar \partial (\partial g g^{-1}) -
 \epsilon_R \partial (g^{-1}\bar\partial g)]}

Defining the traceless, anti-hermitian matrix-valued currents
\eqn\j{\eqalign{&
 J=\partial g g^{-1}\cr&\bar J =-g^{-1}\bar \partial g}}  eq.
\dels\ implies that for $g$ which solves the equations of motion

\eqn\cons{\bar \partial J = \partial \bar J =0} This expresses the
$ \hat{SU(2)}_L \times \hat{SU(2)}_R$ symmetry of the WZW model.

Putting an orientifold into the group manifold amounts to modding out
by the transformation $R\Omega$, where $R$ is
 a $Z_2$ involution of the group manifold and $\Omega$ is the reversal
 of world sheet orientation. We will choose \huss\ the orientifold position such
that it identifies configurations related by
\eqn\orient{g(z,\bar
z)\to g^{-1}(-\bar z,-z)} From the definitions of the currents it
follows that under the
 transformation \orient
\eqn\orcur{\eqalign{&J(z,\bar z)\to -\bar J(-\bar z,-z)\cr&\bar
J(z,\bar z)\to - J(-\bar z,-z)}}

\subsec{Antipodal Mirror Branes}

 Put also a pointlike $D$ brane on the group manifold. Choose  first
its location at $g=h$
 with $h$ such that
its mirror image under the transformation $R$ of the orientifold
is at the antipodal point, namely, $h^{-1}=-h$. For definiteness
let us fix
\eqn\h{h=exp i( {\pi \over 2} \sigma _3)}
 Consider then the open string connecting the $D$ brane
at $h$ to its mirror image at $-h$. (see fig.1)

\vskip 1cm \centerline{\epsfxsize=60mm\epsfbox{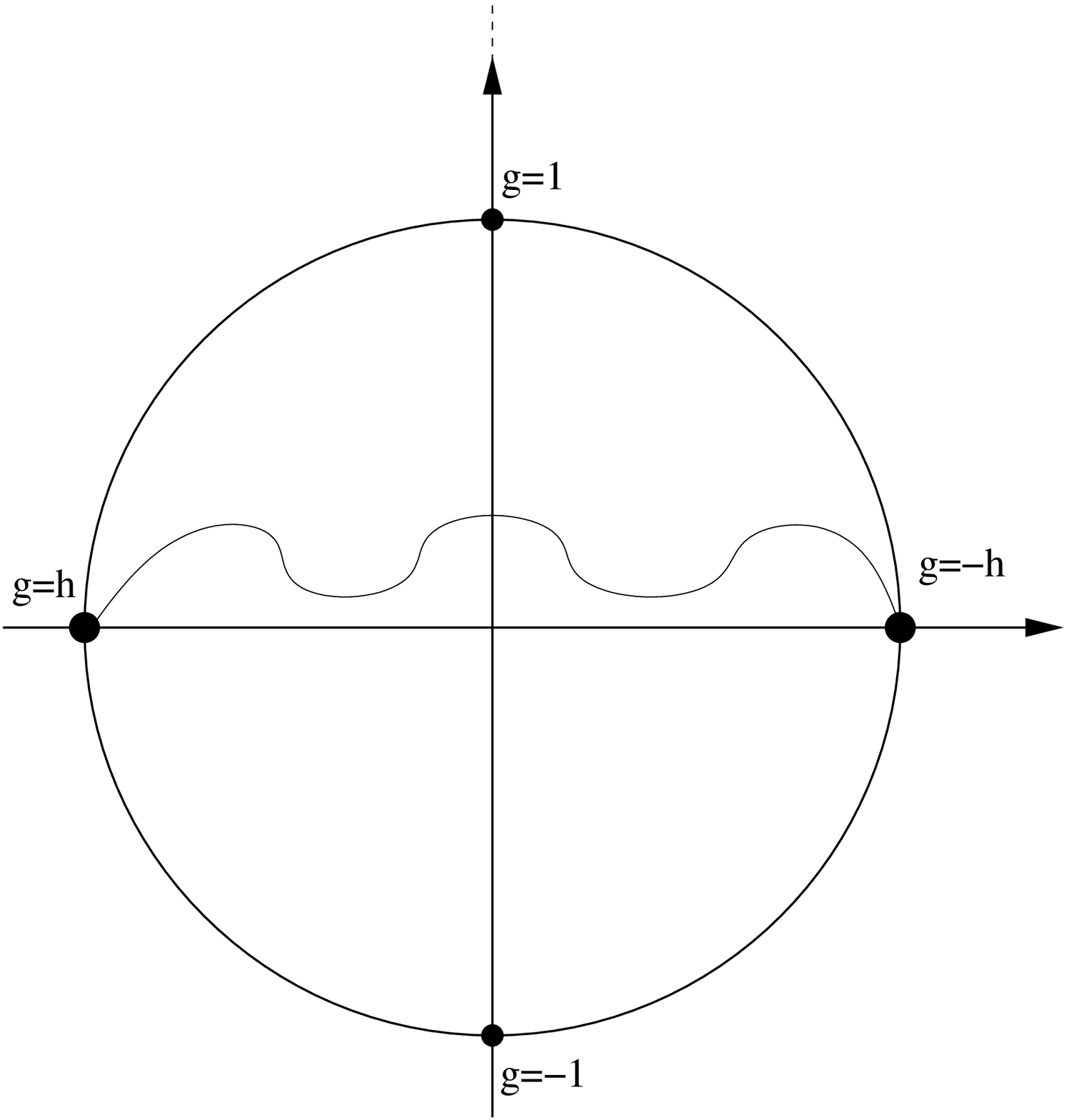}}

\centerline{Fig.\ 1}
\vskip .5cm

Realize the world sheet of this string as the upper half plane
with the boundary conditions
 \eqn\bcl{ g=h=exp i( {\pi \over
2}\sigma_3)} on the negative real axis and
\eqn\bcr{ g=h^{-1}=exp-
i( {\pi \over 2}\sigma_3)} on the positive real axis. The boundary
conditions \bcl\ imply that on the real axis
$(\partial+\bar{\partial})g=0$. In terms of the currents this
gives the boundary conditions \eqn\jb{J-h\bar J h^{-1}=0} on the
real axis.

The infinitesimal variation \var\ is consistent with the boundary
conditions only if \eqn\convar{\epsilon_L = h\epsilon_R h^{-1}} on
the boundary. The induced variation of the action \dels\ can be
written as, \eqn\delos{\eqalign{& \delta S \sim \int d^2 z
Tr(\epsilon_L \bar{ \partial} J+ \epsilon_R
\partial \bar J)\cr&=\int d^2 z Tr{1\over 2}[(\epsilon_L +
h\epsilon_R h^{-1}) (\bar{\partial}J+\partial(h\bar J h^{-1}) +
(\epsilon_L - h\epsilon_R h^{-1}) (\bar{\partial}J-\partial(h\bar
J h^{-1})]}}

The combination $\epsilon_L + h\epsilon_R h^{-1}$ in \delos\ is
arbitrary while the combination $\epsilon_L - h\epsilon_R h^{-1}$
is forced to vanish on the boundary by \convar. Therefore on the
equations of motion the conservation law
\eqn\cono{\bar{\partial}J+\partial(h\bar J h^{-1})=0} holds
everywhere, while the remaining relations,
$\bar{\partial}J-\partial(h\bar J h^{-1})=0$, may be violated on
the boundary. This reflects the fact that the two pointlike branes
at $g=\pm h$ break the
 $ \hat{SU(2)}_L \times \hat{SU(2)}_R$ symmetry of the model in the open
string sector, down to a single $\hat{SU(2)}$ generated by the
current $(J, h\bar J h^{-1})$. It further follows from \cono\ that
everywhere on the upper half-plane, except perhaps at the origin,
\eqn\conon{\bar{\partial}(z^nJ)+\partial(\bar{ z}^n(h\bar J
h^{-1}))=0} for any integer $n$. This is because in the bulk of
the half-plane each of the two terms in \conon\ vanishes
separately, on the boundary $z=\bar z$ and one gets back \cono.

It follows that the quantities, \eqn\modes{\int [z^n J dz -
\bar{z}^n h\bar J h^{-1} d\bar{z}]} where the integral is taken on
any closed path , are zero.  On the boundary the integrand in
\modes\ vanishes identically due to \jb. Therefore if we choose
the contour of integration as the arc $z=re^{i\theta},\pi\ge
\theta \ge 0$ we get that the modes
\eqn\modarc{\eqalign{&J_n={1\over{2\pi i}}\int [z^n J dz -
\bar{z}^n h\bar J h^{-1} d\bar{z}]\cr&=
{1\over{2\pi}}r^{(n+1)}\int_0^{\pi}d\theta[
e^{i(n+1)\theta}J(r,\theta)+ e^{-i(n+1)\theta}h\bar{
J}(r,\theta)h^{-1}] }} are constants of motion. They satisfy the
affine Lie algebra relations for $\hat{SU(2)}_k$.

According to \orcur\ the orientifold identification
takes $J(r,\theta)$ in \modes\
 to  $-\bar {J}(r,\pi - \theta)$. So this identification acts on the modes as,
\eqn\ormod{J_n \to( -1)^n hJ_n h^{-1}}

This is also consistent with the role of the modes $J_n$ as the
Laurent coefficients for $J(z)$. In the bulk of the upper half
plane $J$ is a holomorphic function and $\bar J$ is
antiholomorphic. The boundary conditions \jb\ allow one to extend
$J$ into an holomorphic function on the whole plane defining
\eqn\ext{J(z,\bar z)= h\bar J(\bar z,z) h^{-1}} for $z$ in the
lower half plane. $J(z)$ so defined on both halves of the complex
plane can be Laurent expanded as \eqn\laur{J(z)=\Sigma_{n\in Z}
{J_n\over{z^{n+1}}}} In terms of the extended $J(z)$, the two
terms integrated on the half circle in eq. \modarc\ combine
together to give the integral of $z^n J(z)$ along the full circle.
This means that $J_n$ defined in eq. \modarc\ are the same Laurent
coefficients defined in \laur. The orientifold identification
\orcur\ reads in terms of the extended $J(z)$ \eqn\exor{J(z)\to -h
J(-z)h^{-1}} This immediately implies the action \ormod\ on the
Laurent coefficients.

 Denote $J_n={i\over 2}\Sigma_{a=1}^3 J^{a}_n
\sigma_a$. We have for the orientifold action \eqn\ormot{J^3_n \to
(-1)^n J^3_n} \eqn\ormopm{ J^{1,2}_n \to-(-1)^n J^{1,2}_n} In
particular for the global $SU(2)$ generators, \eqn\orglot{J^3_0
\to J^3_0} \eqn\orglopm{J^\pm _0 \to -J^\pm _0}
\lref\as{ A.~Y.~Alekseev and V.~Schomerus,
Phys.\ Rev.\ D {\bf 60}, 061901 (1999) [arXiv:hep-th/9812193].
}
\lref\cardy{ J.~L.~Cardy,
Nucl.\ Phys.\ B {\bf 324}, 581 (1989).
}

The two branes at $g=\pm h$ preserve, as seen above, the
$\hat{SU(2)}$ symmetry generated by \modes. These two points
correspond to the two trivial conjugacy classes of $SU(2)$, the
points $\pm1$, shifted by the element $h$. According to Alekseev
and Schomerus \as, there are altogether $k+1$ (shifted) conjugacy
classes which are allowed to inhabit branes which preserve this
$SU(2)$ symmetry. Each of these classes corresponds to one of the
$k+1$ primary fields of $SU(2)_k$. In particular the class at $ h$
corresponds to the primary field of spin $0$, while that at $-h$
corresponds to the primary field of spin ${k\over 2}$.
 This correspondence, due to Cardy \cardy, implies that the open
strings stretched between two branes, belong to the
representations of the chiral algebra which appear in the fusion
of the primary fields corresponding to these branes. In our case
then, the strings stretched between the brane at $h$ and that at
$-h$ belong to the representation of the $\hat{SU(2)}$ group
generated by \modes\ with spin ${k\over 2}$. The lightest of those
strings form a degenerate multiplet of $k+1$ members transforming
in the spin ${k \over 2}$ representation under the global $SU(2)$
generated by $J_0$ of \modes. Let $|m>$ be the state
 of such a light string which satisfies $J^3_0 |m>=m|m>$.
Here, ${k \over 2}\ge m \ge -{k \over 2}$ and ${k \over 2}-m$ is
an integer. Let $V_m$ be the vertex operator to emit an open
string in the state $|m>$. The dimension of each of the  $V_m$
 operators is ${1 \over
{k+2}}[{k \over 2}({k\over 2}+1)] = {k\over 4}$.

It follows from eq. \orglot\ that the value of $m$, i.e. the
eigenvalue of $J^3_0$, is preserved by the orientifold
identification. The state $|m>$  must then be taken by this
identification to itself up to a complex coefficient. Due to the
$Z_2$ nature of this identification, it should either be taken to
itself or to minus itself. Suppose that the orientifold
identification does not affect the string in state $|m>$, namely,
\eqn\mtm{ |m> \to|m>}  By standard spin algebra,
 $|m-1> \sim J^-_0 |m>$. Since by \orglopm\ $ J^-_0$ is taken to
minus itself one concludes that \eqn\flip{|m-1> \to -|m-1>} If
the sign of the orientifold is such that it preserves the state
$|{k\over 2}>$, then all the open strings of types
$|{k\over 2}-2n>$ will be preserved as well, while the string states
 $|{k\over 2}-2n+1>$ will be projected out by the
orientifold. In particular, in such a case, the open string
generated by $V_{-{k\over 2}}$ is preserved for $k$ even but
projected out for $k$ odd.

When there are $N$ D branes, rather than one, each
 open string state connecting the $i$th brane to the  mirror
$j$th brane, carries a pair of Chan-Paton indices $(i,j)$. The
orientifold identification for $|(i,j);{k\over 2}>$ should read
now \foot{Apparently the orientifold transformation could also act
on the Chan-Paton indices \pol. Such a $Z_2$ type action has the
general form of $|(i,j)>\to \pm P_{jj'}P^{-1}_{ii'}|(j',i')>$ with
$P$ a unitary matrix. However, the $U(N)\times U(N)$ gauge
symmetry, present in the system before the introduction of the
orientifold, enables one to choose a Chan-Paton basis for which
$P=1$.} \eqn\oriencol{|(i,j);{k\over 2}> \to \pm |(j,i);{k\over
2}>} For the plus sign in \oriencol, the arguments above imply
that $|(i,j);{k\over 2}-2n>$ survives the orientifold projection
only in symmetric combinations of the Chan-Paton indices , while
for $|(i,j);{k\over 2}-2n+1>$ only antisymmetric Chan-Paton
combinations survive. The opposite assignments occur for the minus
sign in \oriencol.

 \subsec{Non-antipodal Branes}

Consider moving the $D$ brane away from the equator putting it at
$g=h$ with \eqn\ha{h=expi({(\pi-\alpha) \over 2}\sigma_3)} (see fig.2)

\vskip 1cm \centerline{\epsfxsize=60mm\epsfbox{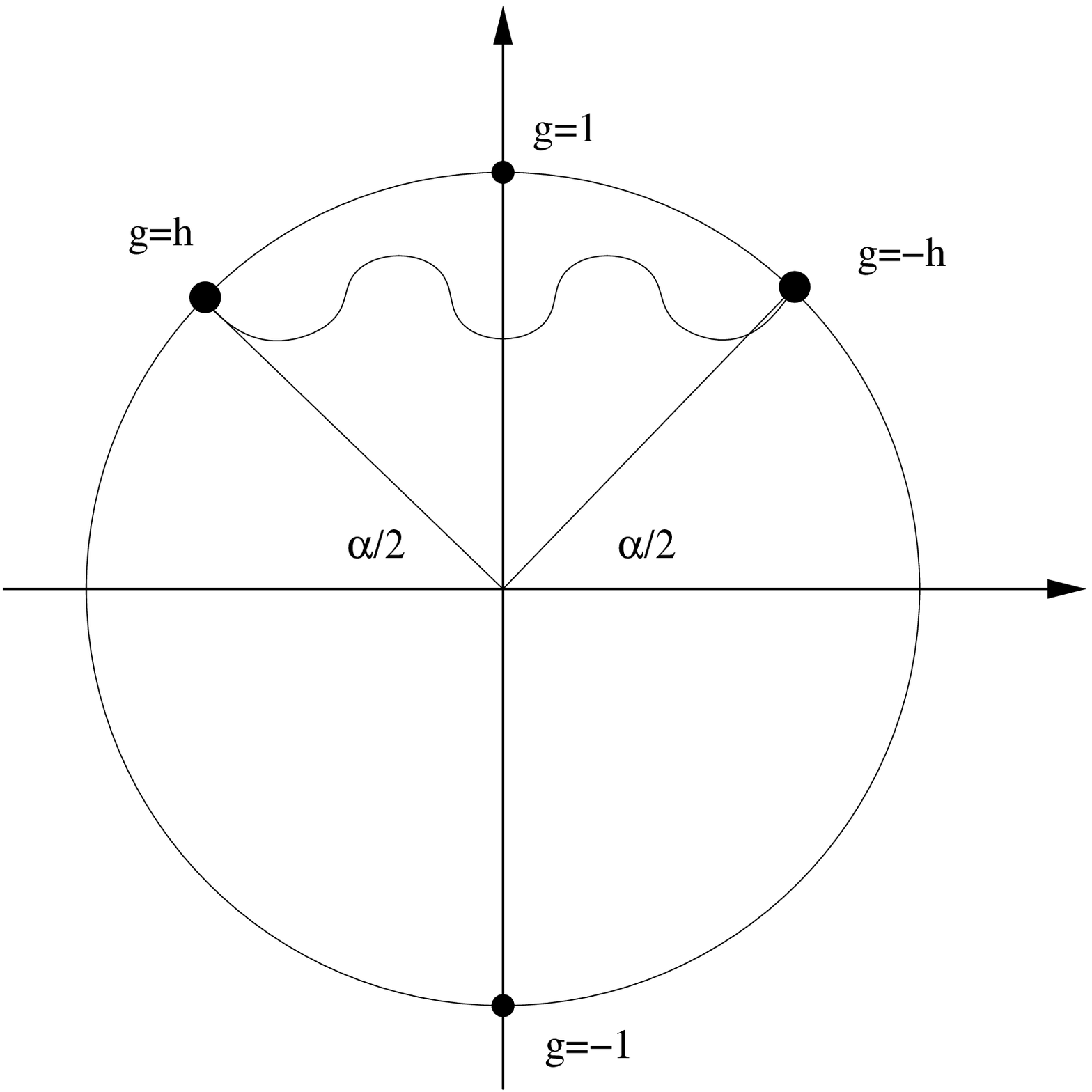}}

 \centerline{Fig.\ 2}
 \vskip .5cm

 The
boundary conditions for an open string connecting this brane to
its mirror brane at $h^{-1}$ are again \eqn\bcla{ g=h=exp i(
{(\pi-\alpha) \over 2}\sigma_3)} on the negative real axis and
\eqn\bcra{ g=h^{-1}=exp(- i( {(\pi-\alpha) \over 2}\sigma_3))= -h
e^{i\alpha \sigma_3}} on the positive real axis. For the currents
, in analogy with \jb, we have \eqn\jbn{J-h\bar J h^{-1}=0} on the
negative real axis and \eqn\jbp {J-h^{-1}\bar J h=0} on the
positive axis.\foot{Choosing for the boundary condition of \jbp\ a
 group element $h$ different than that chosen for \jbn\ is not
 consistent with the orientifold action.}  Now the infinitesimal transformation
\var\ is consistent with the boundary conditions only if the
parameters $\epsilon_{L,R}$ satisfy, \eqn\convaran{\epsilon_L=
h\epsilon_R h^{-1}} on the negative real axis, and
\eqn\convarap{\epsilon_L =h^{-1}\epsilon_L h} on the positive real
axis. In terms of the components of $\epsilon_{L,R}$ defined as
 $\epsilon_{L,R}=i(\epsilon_{L,R}^3 \sigma_3 +\epsilon_{L,R}^-
 \sigma_+ +\epsilon_{L,R}^+ \sigma_-)$,
(where $\sigma_{\pm}={1\over 2}(\sigma_1\pm i\sigma_2)$ )
 the constraints on the $\epsilon$  parameters read,
\eqn\comvan{\eqalign{&\epsilon_L^3=\epsilon_R^3\cr&\epsilon_L^+
=-e^{i\alpha}\epsilon_R^+\cr&\epsilon_L^-
=-e^{-i\alpha}\epsilon_R^-}} on the negative axis, and
\eqn\comvap{\eqalign{&\epsilon_L^3=\epsilon_R^3\cr&\epsilon_L^+
=-e^{-i\alpha}\epsilon_R^+\cr&\epsilon_L^-
=-e^{i\alpha}\epsilon_R^-}} on the positive axis.

Unlike eq. \convar\ here the constraints on the parameters on the
positive real axis differ from those on the negative axis.
Proceeding as in \delos\ shows that the boundary conditions on the
left preserve a different $SU(2)$ subgroup than that preserved by
the boundary conditions on the right. Altogether the two mirror
branes (even without the orientifold) preserve no common $SU(2)$.
Still one can define \eqn\eptil{\eqalign{&\tilde{\epsilon}_L^+=
z^{-{\alpha\over{\pi}}}e^{i{\alpha\over 2}}\epsilon_L^+\cr&
\tilde{\epsilon}_L^-=z^{{\alpha\over{\pi}}}e^{-i{\alpha\over
2}}\epsilon_L^-}} Similarly
\eqn\eptir{\eqalign{&\tilde{\epsilon}_R^+=
\bar{z}^{-{\alpha\over{\pi}}}e^{-i{\alpha\over 2}}\epsilon_R^+\cr&
\tilde{\epsilon}_R^-=\bar{z}^{{\alpha\over{\pi}}}e^{i{\alpha\over
2}} \epsilon_R^-}} Here the cut in the fractional powers of $z$ is
chosen along the negative real axis. In terms of the
$\tilde{\epsilon}$ parameters the boundary conditions constraints
are the same on the positive and negative axis,
\eqn\comtil{\eqalign{&\epsilon_R^3=\epsilon_L^3\cr&
\tilde{\epsilon}_R^+
=-\tilde{\epsilon}_L^+\cr&\tilde{\epsilon}_R^-
=-\tilde{\epsilon}_L^-}} As in \delos\ we can then identify the
currents preserved by \bcla, \bcra. Defining the current
components $J={i\over 2}(J^3 \sigma_3 +J^-
 \sigma_+ +J^+ \sigma_-)$,
the variation \dels\ is
\eqn\deloas{\eqalign{
\delta S \sim \int d^2& z (\epsilon_L^3 \bar{ \partial} J^3 +
\epsilon_L^+\bar{ \partial}J^- +\epsilon_L^-\bar{ \partial}J^+ +\cr&
 \epsilon_R^3  \partial \bar{ J}^3 +
\epsilon_R^+ \partial\bar{J}^- +\epsilon_R^- \partial\bar{J}^+)}}
which, in terms of the $\tilde{\epsilon}$ parameters, reads,
\eqn\tildel{\eqalign{& \delta S \sim \int d^2 z (\epsilon_L^3
\bar{ \partial} J^3 + \tilde{\epsilon}_L^+\bar{
\partial}(e^{-i{\alpha\over 2}} z^{{\alpha\over \pi}}J^-)
+\tilde{\epsilon}_L^-\bar{ \partial}(e^{i{\alpha\over 2}}
z^{-{\alpha \over \pi}}J^+) +\cr&
 \epsilon_R^3  \partial \bar{ J}^3 +
\tilde{\epsilon}_R^+ \partial(e^{i{\alpha\over 2}}
\bar{z}^{{\alpha \over \pi}}\bar{J}^-)
 +\tilde{\epsilon}_R^- \partial(e^{-i{\alpha\over 2}}
\bar{z}^{-{\alpha\over \pi}}\bar{J}^+)}} This can be split  as
\eqn\spl{\delta S=\delta S_1 + \delta S_2} where
\eqn\delone{\eqalign{\delta S_1\sim \int d^2 z&[ (\epsilon_L^3+
\epsilon_R^3)(\bar{ \partial} J^3 + \partial \bar{ J}^3)\cr&+
(\tilde{\epsilon}_L^+-\tilde{\epsilon}_R^+) (\bar{
\partial}(e^{-i{\alpha\over 2}}z^{{\alpha\over \pi}}J^-)-
\partial(e^{i{\alpha\over 2}}\bar{z}^{{\alpha \over \pi}}\bar{J}^-))\cr&+
(\tilde{\epsilon}_L^--\tilde{\epsilon}_R^-) (\bar{
\partial}(e^{i{\alpha\over 2}}z^{-{\alpha \over \pi}}J^+)-
\partial(e^{-i{\alpha\over 2}}\bar{z}^{-{\alpha\over \pi}}\bar{J}^+))]}}

\eqn\deltwo{\eqalign{\delta S_2\sim \int d^2 z&[ (\epsilon_L^3-
\epsilon_R^3)(\bar{ \partial} J^3 - \partial \bar{ J}^3)\cr&+
(\tilde{\epsilon}_L^++\tilde{\epsilon}_R^+) (\bar{
\partial}(e^{-i{\alpha\over 2}}z^{{\alpha\over \pi}}J^-)+
\partial(e^{i{\alpha\over 2}}\bar{z}^{{\alpha \over \pi}}\bar{J}^-))\cr&+
(\tilde{\epsilon}_L^-+\tilde{\epsilon}_R^-) (\bar{
\partial}(e^{i{\alpha\over 2}}z^{-{\alpha \over \pi}}J^+)+
\partial(e^{-i{\alpha\over 2}}\bar{z}^{-{\alpha\over \pi}}\bar{J}^+))]}}

In \delone\ the combinations of the $\tilde{\epsilon}$ parameters
are not constrained by the boundary conditions. The equations of
motion then imply, \eqn\cona{\eqalign{&\bar{\partial}J^3+\partial
\bar {J}^3=0\cr& \bar{\partial}\tilde{J}^+-\partial \tilde{\bar
J}^+=0\cr& \bar{\partial}\tilde{J}^--\partial \tilde{\bar J}^-=0}}
 where
\eqn\tij{\eqalign{& \tilde {J}^+=e^{i{\alpha\over 2}}z^{-{\alpha
\over \pi}}J^+\cr& \tilde {J}^-=e^{-i{\alpha\over
2}}z^{{\alpha\over \pi}}J^-}} and \eqn\tijb{\eqalign{&
\tilde{\bar{J}}^+= e^{-i{\alpha\over 2}}\bar{z}^{-{\alpha\over
\pi}}\bar{J}^+\cr& \tilde{\bar{J}}^-= e^{i{\alpha\over
2}}\bar{z}^{{\alpha \over \pi}}\bar{J}^-}} Note that \jbn\ and
\jbp\ imply the boundary conditions
\eqn\tjb{\eqalign{&J^3-\bar{J}^3=0\cr&\tilde{J}^++\tilde{\bar{J}}^+=0\cr&
\tilde{J}^-+\tilde{\bar{J}}^-=0}} everywhere on the real axis. On
the other hand the transformation parameters in $\delta S_2$ are
constrained to vanish on the boundary, hence the corresponding
current combinations are not conserved on the boundary. Again for
any integer $n$ we have  a conservation analogous to \cona,
\eqn\conan{\eqalign{&\bar{\partial}(z^n J^3) +\partial(\bar{z}^n
\bar {J}^3)=0\cr& \bar{\partial}(z^n \tilde{J}^+)
-\partial(\bar{z}^n \tilde{\bar J}^+)=0\cr&
\bar{\partial}(z^n\tilde{J}^-) -\partial(\bar{z}^n \tilde{\bar
J}^-)=0}} since, as before, in the bulk of the half-plane each
term for each component of \cona\ vanishes separately, while on
the boundary $z=\bar{z}$ so \conan\ reduces to \cona.

As a result of  \conan\  the modes
\eqn\moth{\eqalign{J^3_n=&{1\over{2\pi i}}\int[z^n J^3 dz
 - \bar{z}^n \bar{J}^3 d\bar{z}]=\cr&{1\over{2\pi }} r^{n+1}\int_0^{\pi}
d\theta
[e^{i(n+1)\theta}J^3(r,\theta)+e^{-i(n+1)\theta}\bar{J}^3(r,\theta)]}}

\eqn\mop{\eqalign{\tilde{J}^+_n&={1\over{2\pi
i}}\int[z^n\tilde{J}^+ dz
 +\bar{z}^n \tilde{\bar{J}}^+ d\bar{z}]\cr&={1\over{2\pi i}}\int[e^{i{\alpha \over 2}}
z^{n-{\alpha \over \pi}}J^+ dz +e^{-i{\alpha \over 2}}\bar
{z}^{n-{\alpha \over \pi}}\bar{J}^+d\bar{z}]\cr& ={1\over{2\pi
}}r^{n+1-{\alpha \over \pi}}\int_0^{\pi} d\theta[ e^{i{\alpha
\over 2}}e^{i(n+1-{\alpha \over \pi})\theta}
J^+(r,\theta)-e^{-i{\alpha \over 2}}e^{-i(n+1-{\alpha \over
\pi})\theta} \bar{J}^+(r,\theta)]}}
\eqn\mom{\eqalign{\tilde{J}^-_n&={1\over{2\pi
i}}\int[z^n\tilde{J}^- dz
 +\bar{z}^n \tilde{\bar{J}}^- d\bar{z}]\cr&={1\over{2\pi i}}\int[e^{-i{\alpha \over 2}}
z^{n+{\alpha \over \pi}}J^- dz +e^{i{\alpha \over 2}}\bar
{z}^{n+{\alpha \over \pi}}\bar{J}^-d\bar{z}]\cr& = {1\over{2\pi
}}r^{n+1-{\alpha \over \pi}}\int_0^{\pi} d\theta [e^{-i{\alpha
\over 2}}e^{i(n+1+{\alpha \over \pi})\theta}
J^-(r,\theta)-e^{i{\alpha \over 2}}e^{-i(n+1+{\alpha \over
\pi})\theta} \bar{J}^-(r,\theta)]}} are conserved.   These modes
do not generate the standard $\hat{SU(2)}$ algebra but rather a
spectrally flowed version of it. Redefining
\eqn\flth{\tilde{J}^3_n=J^3_n-{k\alpha \over {2 \pi}}\delta_{n,0}}
the modified  generators $\tilde{J}^3_n, \tilde{J}^+_n$ and $
\tilde{J}^-_n$ generate a standard affine $\hat{SU(2)_k}$ algebra.
The Sugawara Virasoro generator $\tilde {L}_0$ corresponding to
these generators is related to the actual $L_0$ operator of our
model as \eqn\movir{L_0=\tilde{L}_0 + {\alpha \over
{\pi}}\tilde{J}^3_0 + {k \alpha^2 \over {4\pi^2}}}

In the previous subsection we had the $k+1$ vertex operators
$V_m$, all of dimension ${k \over 4}$, creating open strings
connecting the brane at $h$ to its mirror image at $h^{-1}$ for
the case $\alpha=0$. Turning on $\alpha$ continuously, these
operators remain in spin ${k\over 2}$ representation of the
modified $\hat{SU(2)}$ generated by the $\tilde{J}_n$. Their
$\tilde{L}_0$ dimension is ${k\over 4}$. The actual $L_0$
dimension is then, by \movir\ \eqn\dimv{h_m={k\over 4}+{\alpha
\over {\pi}}m + {k \alpha^2 \over {4 \pi^2}}}

The orientifold  identification \orcur, takes $J^a (r,\theta)$ in
\mop\ and \mom\ into $-\bar {J}^a ( r,\pi - \theta)$. Applying
this transformation to \mop, \mom, we see that the modes
$\tilde{J}_n$ transform under this identification, for general
$\alpha$, in the same manner as in eq. \ormot, \ormopm, for
$\alpha=0$, namely \eqn\ortt{\tilde{J}^3_n\to (-1)^n
\tilde{J}^3_n} \eqn\ortpm{\tilde{J}^{\pm}_n\to -(-1)^n
\tilde{J}^{\pm}_n}

 As in previous subsection, here also in the case of non antipodal branes,
  this action of the orientifold on $\tilde{J}_n$ can be deduced
 from their role as Laurent coefficients. Again the
functions $\tilde {J}^3(z), \tilde{J}^\pm(z)$ can be extended into
the lower half plane defining \eqn\nexth{\tilde {J}^3(z)=\tilde
{\bar J}^3(\bar z)} \eqn\nexpm{\tilde {J}^\pm (z)=-\tilde {\bar
J}^\pm (\bar z)} for $z$ in the lower half plane. The function is
analyitc on the real axis due to the boundary conditions \tjb.
Equations \moth, \mop\ and \mom\ identify then $\tilde {J}^3_n,
\tilde{J}^\pm_n$ as the Laurent coefficients of $\tilde {J}^3(z),
\tilde{J}^\pm(z)$, \eqn\tlaurth{\tilde {J}^3(z)=\Sigma{\tilde
{J}^3_n\over {z^{n+1}}}} \eqn\tlaurpm{\tilde {J}^\pm
(z)=\Sigma{{\tilde {J}^\pm _n}\over {z^{n+1}}}} The orientifold
identification \orcur\ takes the extended function $\tilde
{J}^3(z)$ into $-\tilde {J}^3(-z)$. Eq. \tlaurth\ implies then
\ortt\ on the Laurent coefficients. As to $\tilde{J}^+(z,\bar z)$,
by its definition \tij\ it is taken by the orientifold action as
\eqn\tort{\tilde{J}^+(z,\bar z)\to -e^{i{\alpha\over
2}}z^{-{\alpha\over\pi}}\bar{J}^+(-\bar z,-z)} According to eq.
\tijb\ $\bar{J}^+(-\bar z,-z)=e^{i{\alpha\over
2}}(-z)^{\alpha\over\pi}\tilde{\bar J}^+(-\bar z,-z)$. For $z$ in
the upper half plane, to avoid the cut along the negative axis,
 $(-z)^{\alpha\over\pi}$ should be read as
$(e^{-i\pi}z)^{\alpha\over\pi}$. Substituting this into \tort\
gives the simple transformation \eqn\tiltort{\tilde{J}^+(z,\bar
z)\to -\tilde{J}^+(-\bar z,-z)} In terms of the extended
hlomorphic $\tilde{J}^+(z)$ defined in \nexpm, this is
\eqn\tilhol{\tilde{J}^+(z)\to \tilde{J}^+(-z)} The same
orientifold action is found for $\tilde{J}^-(z)$. The Laurent
expansion \tlaurpm\ implies then \ortpm\ for the orientifold
action on $\tilde {J}^\pm _n$.

In particular for $n=0$ we have \eqn\orgtt{\tilde{J}^3_0\to
\tilde{J}^3_0} \eqn\orgtpm{\tilde{J}^{\pm}_0\to -
\tilde{J}^{\pm}_0} The same argument as in previous subsection
implies then that if the orientifold is chosen such that its
identification takes the vertex operator $V_{k\over 2}$ to itself
then the vertex operators $V_{{k\over 2}-2n}$ for integer $n$ will
survive the orientifold projection while the operators $V_{{k\over
2}-2n+1}$ will be projected out. More generally, for $N$ branes,
the surviving operators of the form \eqn\vev{V_{{k\over
2}-2n}^{(i,j)}} which emits the open string in state
$|(i,j);{k\over 2}-2n>$ should be symmetric in the Chan-Paton
indices while those of the form
 \eqn\vod{V_{{k\over 2}-2n+1}^{(i,j)}}
 are antisymmetric.
 This is the same behavior as
in previous subsection, now for a general value of $\alpha$.

\subsec{Coincident Branes}

 The main conclusion of previous discussion is that $V_{k\over 2}$ and
$V_{-{k\over 2}}$ behave the same way under the orientifold
projection for even $k$ and in opposite ways for odd $k$. In this
subsection we check this conclusion for the limiting cases
$\alpha=\pm \pi$, where it can be looked at from a different point
of view. By eq. \flth\ with $\alpha=\pi$ the $J^3$ charge of $V_m$
is ${k\over 2}+m$. The dimension of $V_m$ is also, according to
eq. \dimv, ${k\over 2}+m$. The operator $V_{-{k\over 2}}$ has then
zero charge and dimension. The other $V_m$ operators with
$m<{k\over 2}$ have positive integral dimensions. For $\alpha=\pi$
the brane and its mirror image coincide at $g=1$. The open strings
created by $V_m$ connect branes sitting at conjugacy classes of
$SU(2)$. Invoking again \as, the conjugacy class at $g=1$
corresponds to the primary field of spin $0$ of the affine
$\hat{SU(2)}$ preserved by the branes. This is the group generated
by the modes $J_n$ defined in \modes, here with $h=1$. The strings
emitted by $V_m$ connect this conjugacy class with itself, hence
\cardy\ they belong to the representation contained in the fusion
of the zero spin primary field with itself, that is to the zero
spin representation. The operator $V_{-{k\over 2}}$ which has zero
dimension and charge should be identified with the primary field
of this representation. The other $V_m$ operators which have
positive integral dimensions and non zero charges are from this
point of view descendants in this zero spin representation.

\lref\difr{P. Di Francesco, P. Mathieu and D. Senechal,
\it{Conformal Field Theory}, NY Springer 1997, p. 684.}%

For $\alpha=-\pi$ the two branes meet at $g=-1$. They preserve the
same $\hat{SU(2)}$ as for $\alpha=\pi$. The conjugacy class at
$g=-1$ correspond to the primary field with the maximal spin
${k\over 2}$. The strings connecting these branes belong to the
representations in the fusion of the primary field with spin
${k\over 2}$ with itself, which is again just the zero spin
representation \difr. Here eq. \flth\ gives the operator $V_m$ the
charge $m-{k \over 2}$ and eq. \dimv\ fixes its dimension to be
${k\over 2}-m$. The primary field of the zero spin representation
should then be the operator $V_{k\over 2}$ which has zero charge
and mass, while the other $V_m$ operators are descendants.

We find then that among the open strings  emitted at $g=1$,
between branes corresponding to the spin zero representation,
$V_{-{k\over 2}}$ is the primary field, while out of those emitted
at $g=-1$ between branes corresponding to spin ${k\over 2}$,
$V_{k\over 2}$ is the primary. In previous section we found that,
for any $\alpha$, if under the orientifold action $V_{k\over 2}
\to V_{k\over 2}$ then under this action $V_{-{k\over 2}} \to
(-1)^k V_{-{k\over 2}}$. We learn then that the action of the
orientifold on strings sitting on the brane corresponding to zero
spin primary field, is related by a factor $(-1)^k$ to its action
on strings which sit on branes corresponding to spin ${k\over 2}$.
This is consistent with the analysis of refs.
\prad,\brunner,\bach,\huss. There it is found that the relative
sign between the annulus and the Mobius strip  on branes in
$SU(2)$ WZW model which correspond to integral spin
representations, differs from that sign for branes corresponding
to half integral spin.

\newsec{Open Strings near NS Branes}

This section is a review of the part of \gor\ relevant for our
subject. Consider a stack of $k>1$ NS-$5$ branes spanning the
hyper-plane $(012345)$ in $10$ dimensional type IIA model,
  at $x^6=x^7=x^8=x^9=0$.  The near
horizon geometry formed by these branes is \chs\
\eqn\nhg{\eqalign{& e^{2 (\Phi-\Phi_0)}= {k\over |x|^2}\cr&
G_{I,J} = e^{2 (\Phi-\Phi_0)}\delta_{I,J} \cr&
G_{\mu\nu}=\eta_{\mu\nu}\cr& H_{IJK}=-\epsilon_ {IJKM}\partial ^M
\Phi}} where $I,J,K,M$ run from $6$ to $10$, $\mu,\nu$ run from
$0$ to $5$ and $|x|^2=(x^6)^2+(x^7)^2+(x^8)^2+(x^9)^2$. This geometry
is shown schematically in fig.3 .

\vskip 1cm \centerline{\epsfxsize=80mm\epsfbox{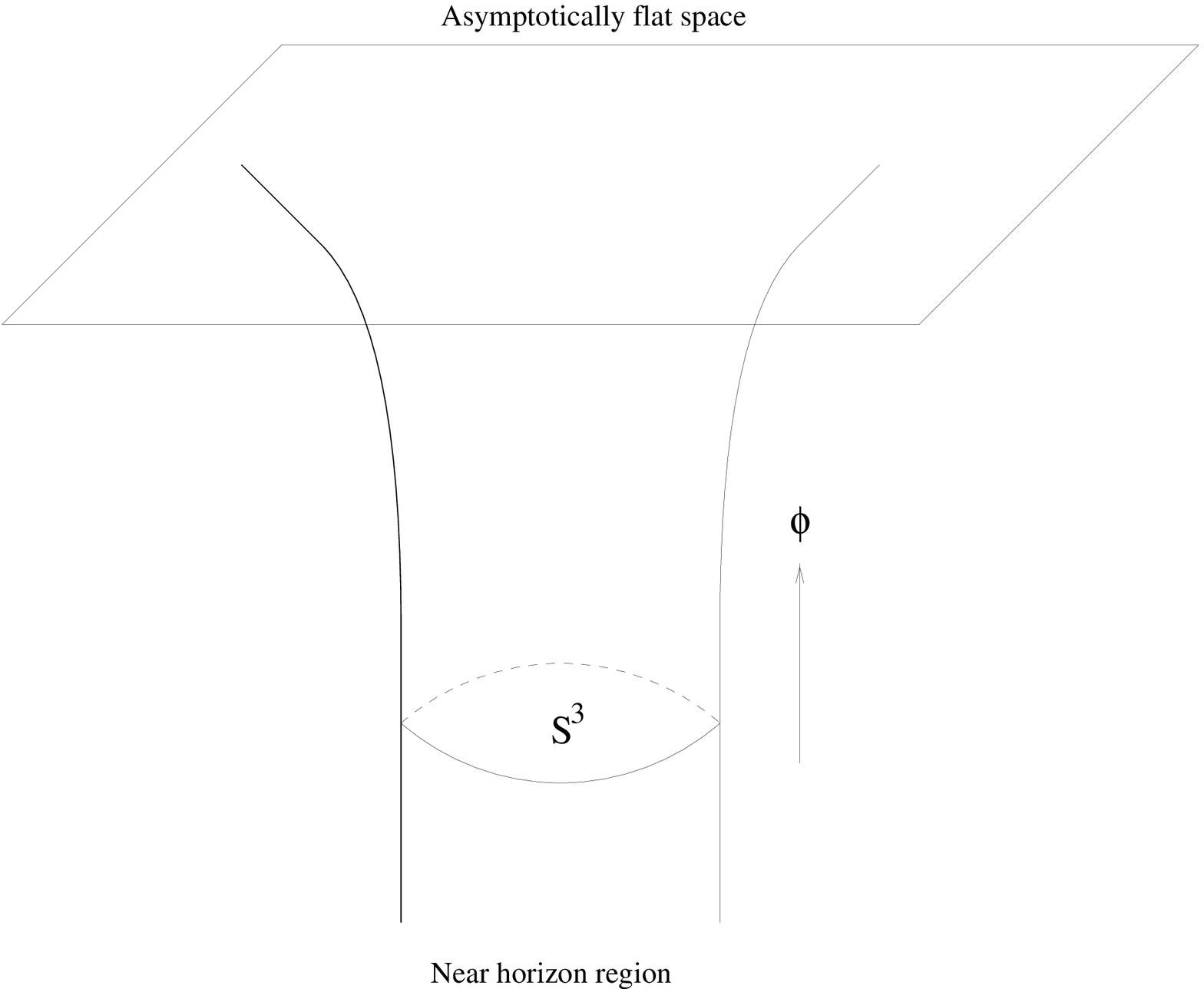}}
\centerline{Fig.\ 3}
 \vskip .5cm

 This
background is described by an exact CFT with the target space
$R^{5,1}\times R_\phi \times SU(2)_k$. Here $R_\phi$ represents
the radial coordinate $|x|$ with a coordinate $\phi$ related to it
by \eqn\phil{\phi={1\over Q}log {|x|^2 \over k}} where
\eqn\q{Q=\sqrt{2\over k}}
 In terms of the coordinate $\phi$ the
dilaton is linear, \eqn\lindil{\Phi=-{Q\over 2} \phi} The
$SU(2)_k$ factor consists of the bosonic angular coordinates
parametrizing $S^3$ in the $4$ dimensional space transverse to the
branes, and the three corresponding fermions
$\chi_1,\chi_2,\chi_3$. A group element $g\in SU(2)$ can be
parametrized as \eqn\g{g={1\over{|x|}}[ x^7 1+i(x^8 \sigma_1
+x^9\sigma_2 + x^6 \sigma_3)]} $g$ is a point in the $\hat{SU(2)}$
bosonic target space of level $k-2$. The three fermions
$\chi^{1,2,3}$ form another representation of $\hat{SU(2)}$ of
level $2$.

Let a stack of $N D4$ branes end on the NS $5$ branes. These are
stretched along the $(0123)$ hyper-plane with their fifth
coordinate in the $(6,7)$ plane forming an angle ${\alpha \over
2}$ with the $x^6$ axis. Let another identical stack end on the NS
branes forming an angle $\pi-{\alpha\over 2}$ with the $x^6$ axis
in the $(6,7)$ plane (see fig.4). On the world sheet of an open
string connecting the first stack of D-branes to the second one
the boundary conditions for $g$ are those of \bcla\ and \bcra.

\vskip 1cm \centerline{\epsfxsize=80mm\epsfbox{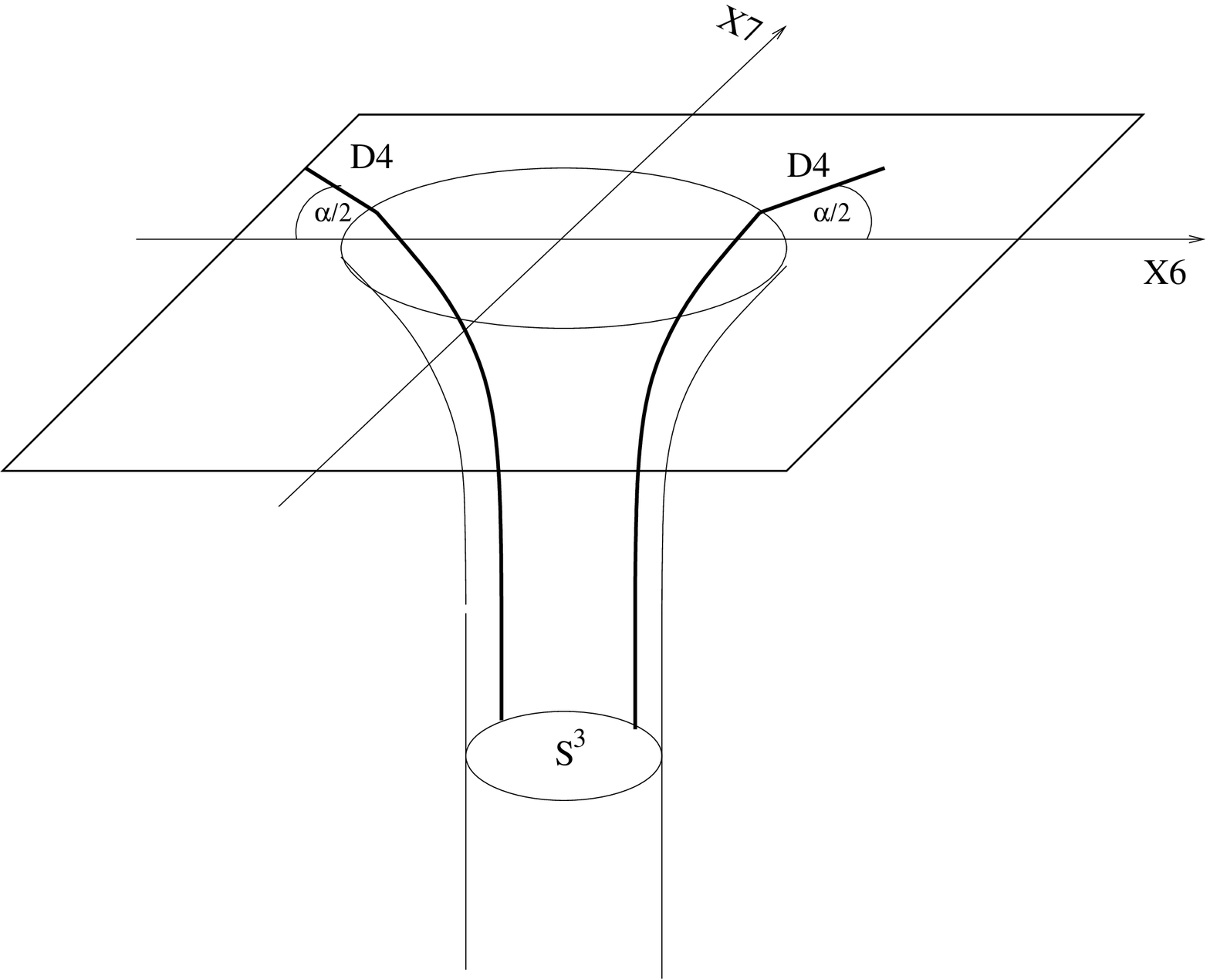}}
\centerline{Fig.\ 4} \vskip .5cm

 Denote
the currents defined in \j\ by $J_B$, the bosonic currents. Their
boundary conditions are those of \jbn,\jbp. By world sheet
supersymmetry the boundary conditions on the  fermions
$\chi^{1,2,3}$ are correlated with those for $J_B$. Denoting
\eqn\chim{\chi=\chi^1 \sigma_1 +\chi^2 \sigma_2+\chi^3 \sigma_3}
the boundary conditions for $\chi$ are \eqn\jfn{\chi\pm h
\bar{\chi}h^{-1}=0} on the negative axis, and \eqn\jfp{\chi\pm
h^{-1} \bar{\chi}h=0} on the positive axis . The signs in \jfn\
and \jfp\ are equal for an open string in the NS sector, they are
opposite for the Ramond sector. Defining \eqn\jf{J_F=[\chi,\chi]}
the fermionic currents $J_F$  satisfy the same boundary conditions
\jbn\ and \jbp\ as those of the bosonic currents $J_B$.

 A NS vertex operator for
emitting a light open string connecting the $i$th $D4$ brane of
the first stack to the $j$th  brane of the second is, in the $-1$
picture, of the form \eqn\vop{V=e^{-\varphi}expi(\Sigma_0^3 k_\mu
x^\mu) e^{\beta \phi}V^{B(i,j)}_{m_1}V^F_{m_2}} where $\varphi$ is
the bosonized susy ghost. The factor $expi(\Sigma_0^3 k_\mu
x^\mu)$ in \vop\ describes the $4$-dimensional motion of the
emitted open string, the factor $e^{\beta \phi}$ is responsible
for its motion along the linear dilaton radial direction. The
factor $V^{B(i,j)}_{m_1}$, acting on the bosonic  $SU(2)$ part, is
the same operator discussed in \vev, \vod\ in the last section. It
creates a string belonging to a spin ${k-2\over 2}$ for the
$SU(2)$ generated by $\tilde{J}_{B0}$ with the value $m_1$ for the
operator $\tilde{J}_{B0}^3$. Its dimension $h_B$ is given by eq.
\dimv, \eqn\dimvb{h_B={k-2 \over 4}+{\alpha \over
\pi}m_1+{(k-2)\alpha^2 \over 4 \pi^2}} $V^F_{m_2}$ is the
corresponding fermionic operator. It creates a state with spin $1$
under the $SU(2)$ generated by $\tilde{J}_{F0}$ with the value
$m_2$ for the operator $\tilde{J}_{F0}^3$. Its dimension $h_F$ is
accordingly, \eqn\dimvf{h_F={2 \over 4}+{\alpha \over
\pi}m_2+{2\alpha^2 \over 4 \pi^2}} The allowed values of the
parameters $k_{\mu}$ and $\beta$ are subject to the mass shell
condition which requires the total dimension of $V$ to be $1$. The
open strings created by $V$ are light in the sense that no string
oscilator is excited in \vop.

As it stands the background \nhg\ is singular, the string coupling
$e^{\Phi}$ diverges when $|x|\to 0$. A regularization which can
avoid this singularity leads to a perturbatively controllable
background. One way for such a regularization is by turning on a
Liouville potential for $\phi$, which shields the region $\phi \to
-\infty$. For the sake of spacetime supersymmetry the $\phi$ field
has to be treated as a component of an  $N=2$  Liouville system.
This system contains also a $U(1)$ generator. This can be chosen
to be the total $J^3$ current, $J^3=J_B^3+J_F^3$. One then
bosonizes the current $J^3$ defining a scalar field $Y$ by
\eqn\y{J^3=2i\sqrt{{k \over 2}}
\partial Y}
treating $\phi$ and $Y$ as components of $N=2$ $2d$
Liouville system.
 Note that the Liouville superpotential required for the regularization
 breaks the $SU(2)$
symmetry, assigning a special role to  $J^3$.

An alternative, equivalent, regularization replaces the linear
dilaton coordinate  $\phi$ and the angular coordinate $Y$,
together with their corresponding fermions,  by the cigar shaped
(super) coset $SL(2,R)/U(1)$ \ogvaf,\sfet,\gkp,\gk. Instead of
shielding the region $\phi \to -\infty$ by a potential wall, it is
cut off the geometry by the tip of the cigar. The resulting CFT
becomes then \foot{The product here is not exactly a direct
product, see \gor\ and references therein for more details.}
\eqn\bcg{R^{1,5}\times SL(2)_k/U(1)\times SU(2)_k/U(1)} the level
of $SL(2)$ is chosen such that the linear dilaton behavior
\lindil\ is reconstructed for large positive $\phi$.
Geometrically, this regularization amounts to splitting the $k$
NS-$5$ branes which generate the configuration, spreading them
with small mutual distances from each other along the $(6,7)$
plane. Here also the $SU(2)$ symmetry is broken by the
regularization.

\lref\ogmal{ J.~M.~Maldacena and H.~Ooguri,
J.\ Math.\ Phys.\  {\bf 42}, 2929 (2001) [arXiv:hep-th/0001053].
}

 In this language, we shall build a vertex operator for emitting a light
open string from a $D$-brane of the first stack to a  brane in the
other stack. Denote by $V'^w_{j,m}$ the  $SL(2)_k$ vertex operator
corresponding to a unitary representation of $SL(2)$ with the
value $-j(j+1)$ for its second Casimir, spectrally flowed $w$
times ($w\in Z$) \ogmal, with $U(1)$ charge $m$. The dimension of
this operator is $h=-{j(j+1)\over k}$. Let $[V'^w_{j,m}]$ denote
the $SL(2)/U(1)$ part of $V'^w_{j,m}$. Accordingly the dimension
of $[V'^w_{j,m}]$ is \eqn\dimp{[h]^w_{j,m}=-{j(j+1)\over
k}+{(m+kw)^2\over k}} Similarly, denote by
$[V^{B(i,j)}_{m_1}V^F_{m_2}]$ the $SU(2)/U(1)$ part of the full
$SU(2)$ operator $V^{B(i,j)}_{m_1}V^F_{m_2}$ of \vop. The total
$J^3_0$ charge of $V^{B(i,j)}_{m_1}V^F_{m_2}$  is, by \flth,
$m_1+m_2+{k\alpha\over 2\pi}$. The dimension of this operator
 is given by eq. \dimv. The dimension $[h]$ of the
$SU(2)$ part $[V^{B(i,j)}_{m_1}V^F_{m_2}]$ is
\eqn\shifdim{[h]={k\over 4}+{\alpha \over {\pi}}(m_1 +m_2) + {k
\alpha^2 \over {4 \pi^2}}-{(m_1+m_2+{k\alpha\over 2\pi})^2\over
k}} In these terms the full vertex operator for emitting a light
open string connecting the two stacks becomes
\eqn\vopr{V=e^{-\varphi}expi(\Sigma_0^3 k_\mu x^\mu)
[V'^w_{j,m_1+m_2+{k\alpha\over 2\pi}}][V^{B(i,j)}_{m_1}V^F_{m_2}]}
 The $J^3_0$ charge in the $SU(2)$ and $SL(2)$
parts of the vertex operator should be the same. They both refer
to the same field $Y$ defined in \y.

 The near horizon geometry
\nhg\ and its cigar regularization are adequate for vertex
operators which are confined to the neighborhood of the NS branes
source. Such are the operators of the form \vopr\ with
$V'^w_{j,m_1+m_2+{k\alpha\over 2\pi}}$ corresponding to a discrete
representation of $SL(2,R)$ whose wave function is exponentially
suppressed  for large $|x|$. In contrast, vertex operators
corresponding to continuous representations, which create long
strings \ogmal, have their wave functions extended into the far
region where the description \nhg\ is not appropriate. We will
then focus attention on discrete representations. For those,  $j$
is a real number. The unitarity bound for $\hat{SL(2,R)}$ models
is $-{1 \over 2}<j<{k-1 \over 2}$ \gk, \ogmal. In discrete
representations of $SL(2)$, if $m$ is the eigenvalue of $J^3$ then
the difference $|m|-j$ is a positive integer.

The mass $M$ of the string emitted by \vopr\ is fixed by the mass
shell condition to satisfy \eqn\mass{{M^2\over 2}={1 \over 2}(
k_0^2-\Sigma_1^3 k_i^2)=-{1 \over 2}-{j(j+1)\over k} +{k\over
4}[1+{\alpha^2\over \pi^2}+4w(w+{\alpha\over \pi})
]+(m_1+m_2)(2w+{\alpha\over \pi})} Here, \eqn\uin{-{k \over 2}\le
m_1+m_2 \le{k \over 2}} by the $SU(2)$ role of $m_{1,2}$, and
\eqn\rin{|m_1+m_2-{k\alpha\over 2\pi}|-j} is a positive integer
because of the $SL(2)$ role of $m_1+m_2+{k\alpha\over 2\pi}$ and
the conditions on discrete representations. For $\alpha=0$  eq.
\mass\ reads \eqn\maz{{kM^2 \over 2}={k\over 2}({k\over
2}-1)-j(j+1)+kw[kw+2(m_1+m_2)]} Subject to the inequality \uin\
and condition \rin\ the mass squared \maz\ is non-negative. This
is consistent with the fact that the background with $\alpha=0$ is
space-time supersymmetric. Choosing in \maz\ $w=0$ and $j={k\over
2}-1$ one gets zero mass. The conditions on discrete
representations fix the value of $m_1+m_2$ to either ${k\over 2}$
or $-{k\over 2}$. So out of the $3(k-2)$ allowed values of
$(m_1,m_2)$ in \vopr, only the two values $(m_1,m_2)=({k-2\over
2},1)$ and $(m_1,m_2)=(-{k-2\over 2},-1)$ correspond to massless
particles. (Remember that the $SU(2)$ symmetry is broken by the
regularization). Note that these two massless states belong to the
representation of the total $SU(2)$ generated by $J_{B0}+J_{F0}$
with spin ${k\over 2}$. For non zero $\alpha$ the masses of these
two states changes according to \mass. For the state with
$m_1+m_2={k\over 2}$ we have $j_+={k\over 2}+{k\alpha\over
2\pi}-1$. Substituting $j_+$ for $j$ in \mass\ we get
\eqn\mp{M^2_+={\alpha \over \pi}}  For the state with
$m_1+m_2=-{k\over 2}$ we have $j_-={k\over 2}-{k\alpha\over
2\pi}-1$. This gives the mass \eqn\mn{M^2_-=-{\alpha \over \pi}}

\newsec{Phase Transition in Presence of an Orientifold}

Let an $O6$ orientifold be stretched along the $(0123457)$ plane,
at the location $x^6=x^8=x^9=0$, in the system described in
previous section. This amounts to gauging the world sheet symmetry
\eqn\orga{x^I(z,\bar z)\to x^I(-\bar z,-z)} for $I=0,1,2,3,4,5,7$
and \eqn\rorga{x^I(z,\bar z)\to -x^I(-\bar z,-z)} for $I=6,8,9$.
To preserve world sheet supersymmetry, the corresponding fermions
are transformed in a correlated way, \eqn\orgaf{\psi^I(z,\bar
z)\to i\bar{\psi}^I(-\bar z,-z)} for $I=0,1,2,3,4,5,7$ and
\eqn\rorgaf{\psi^I(z,\bar z)\to -i\bar{\psi}^I(-\bar z,-z)} for
$I=6,8,9$. \foot{The factor $i$ in \orgaf\ and \rorgaf\ results
from the world sheet spin $1/2$ of the operators $\psi^I$. The
transition from $z,\bar{z}$ to $-\bar{z},-z$ changes the sign of
the single $z$ and $\bar z$ derivative in the fermionic world
sheet action. This has to be restored by these factors of $i$.} In
terms of the coordinates $x^\mu,|x|$ and $g$ in \nhg, the
orientifold action is \eqn\corg{\eqalign{&x^\mu(z,\bar z) \to
x^\mu (-\bar z, -z)\cr&|x|(z, \bar z) \to |x|(-\bar z,-z) \cr&
g(z, \bar z) \to g^{-1}(-\bar z,-z)}} Since the action on $g$ is
identical to that of eq. \orient\ the results of sec. 2 for the
action of the orientifold on the bosonic currents and their modes
apply to our case. By \orcur\ this action on the bosonic current
is \eqn\orcurb{J_B(z,\bar z) \to -\bar{J}_B(-\bar z,-z)} On the
world sheet of the open string which connects the $D4$ branes to
their orientifold images, the modes of the bosonic current
$\tilde{J}^3_{Bn},\tilde{J}^{\pm}_{Bn}$ are defined as in
\moth,\mop,\mom\ and \flth. According to \ortt\ and \ortpm, the
orientifold identifies these modes under the transformation
\eqn\ortb{\tilde{J}^3_{Bn}\to (-1)^n\tilde{J}^3_{Bn}}
\eqn\orpmb{\tilde{J}^{\pm}_{Bn} \to-(-1)^n\tilde{J}^{\pm}_{Bn}} By
 \rorgaf\ the orientifold action on the $\chi$ fermions is,
\eqn\orf{\chi(z,\bar z) \to -i\bar{\chi}(-\bar z,-z)} with matrix
valued $\chi$ defined in \chim. This implies the same type of
action on $J_F$ defined in \jf. \eqn\orjf{J_F(z,\bar z) \to
-\bar{J}_F(-\bar z,-z)} As discussed in the previous section, the
fermionic current $J_F$ has to have the same boundary conditions
as the bosonic current $J_B$. The fermionic current has then the
same type of mode expansion as the bosonic one, whose coefficients
$\tilde{J}_{Fn}$ are defined analogously to \moth, \mop, \mom\ and
\flth, with $k=2$ in \flth. Since the orientifold action on the
local currents is the same on the bosonic and fermionic currents,
it is also the same on the modes. The orientifold action on the
fermionic modes is then also of the form \ortb\ and \orpmb,
\eqn\ortf{\tilde{J}^3_{Fn}\to (-1)^n\tilde{J}^3_{Fn}} \eqn\orpmf
{\tilde{J}^{\pm}_{Fn} \to-(-1)^n\tilde{J}^{\pm}_{Fn}} We get then
the same type of action also for the modes of the total current
$J=J_B+J_F$. For the modes of this current we get again eqs.
\ortt\ and \ortpm, and for $n=0$ eqs. \orgtt\ and \orgtpm, for
their orientifold identification.

 The arguments of sec. 2 can be repeated here to show
that if $V^S_m$ is such a vertex operator which transforms
according to the spin $S$ representation under the $SU(2)$ group
generated by the total currents $\tilde {J}^3_0,\tilde
{J}^{\pm}_0$ with $m$ being the eigenvalue of $\tilde {J}^3_0$,
then the orientifold action changes sign between $V^S_m$ and
$V^S_{m+1}$. In particular, if $V^S_S$ is identified with itself
under the orientifold action, then, for $S$ integer, $V^S_{-S}$ is
also identified with itself and for $S$ half integer $V^S_{-S}$ is
identified with minus itself.

Let us follow the low energy behavior of the theory on the stack
of $N D4$ branes as a function of the parameter $\alpha$. To have
a $4$ dimensional physics assume that these branes connect our
pile of $NS 5$ branes  to another $NS$ brane , call it the $NS'$
brane, sitting in the $(6,7)$ plane at a finite distance from the
origin . The mirror images of these $D$ branes should of course
end at the mirror image of this $NS'$ brane under the orientifold
projection. We change the value of $\alpha$ by moving the $NS'$
brane around in the $(6,7)$ plane. Start with the supersymmetric
case $\alpha=0$. In the absence of the orientifold the two stacks
of $D4$ branes are not identified with each other. We expect at
low energy a $U(N)\times U(N)$ gauge group. We found two types of
massless open strings connecting one stack to the other,
corresponding to $m_1+m_2=\pm{k \over 2}$. The vertex operator
with $m_1+m_2={k \over 2}$ gives rise to a complex massless chiral
superfield $Q$, in the bi-fundamental representation $(\bar N,N)$.
The operator with $m_1+m_2=-{k \over 2}$ gives rise to an
anti-chiral superfield denoted by $\tilde {Q}^*$, in the same
representation. $Q$ in the $(\bar N,N)$ and $\tilde{Q}$ in the
$(N,\bar N)$ representation are two chiral fields with $J^3_0$
charge ${k\over 2}$.

The orientifold identifies the two sets of branes, turning the
gauge group into a single $U(N)$. For even $k$, we found that both
$Q$ and $\tilde {Q}^*$ have the same orientifold behavior. If the
sign of the orientifold is chosen to take the vertex operators
corresponding to $Q$ and $\tilde {Q}^*$ to themselves, their
Chan-Paton symmetric part survives the projection and the
resulting $U(N)$ gauge theory contains two chiral matter
superfields, one in the symmetric representation and the other in
its conjugate. For the other choice of the sign of the orientifold
only the antisymmetric part survives and we get a theory with two
fields, one in the antisymmetric representation of $U(N)$ and the
other in its conjugate. For an odd $k$ we find that the two vertex
operators corresponding to $Q$ and $\tilde{Q}^*$ have opposite
behavior under the orientifold transformation. In that case then
the low energy gauge theory will contain two massless matter
fields, $Q$ in the symmetric representation and $\tilde Q$ in the
conjugate antisymmetric one (or vice versa). This is indeed the
matter content suggested in \lll\ and \tsabar\ for the case $k=1$.

Now rotate the $D4$ branes to get $\alpha>0$. Notice that the
preceding analysis of the orientifold action on $Q$ and $\tilde
{Q}^*$ applies for any value of $\alpha$. By eq. \mp\ the mass
squared $M^2_-$ becomes negative. The scalar component of the
field $\tilde Q$ becomes tachyonic. Suppose the sign of the
orientifold is such that this field is in the symmetric
representation of $U(N)$. From field theory point of view this
means that $\tilde Q$ gets a non zero vacuum expectation value. An
expectation value for a symmetric field breaks the gauge group
$U(N)$ down to $SO(N)$. From a geometrical point of view the
appearance of a tachyon is a sign of an instability of the system
caused by the rotation of the branes into a non zero $\alpha$
position. Presumably the system decays into a more stable position
in which the $D4$ branes detach from the pile of the $NS$ branes
connecting the $NS'$ brane directly to its image, meeting the
orientifold at right angle at a point with $x^7>0$ (see fig.5).

\vskip 1cm \centerline{\epsfxsize=90mm\epsfbox{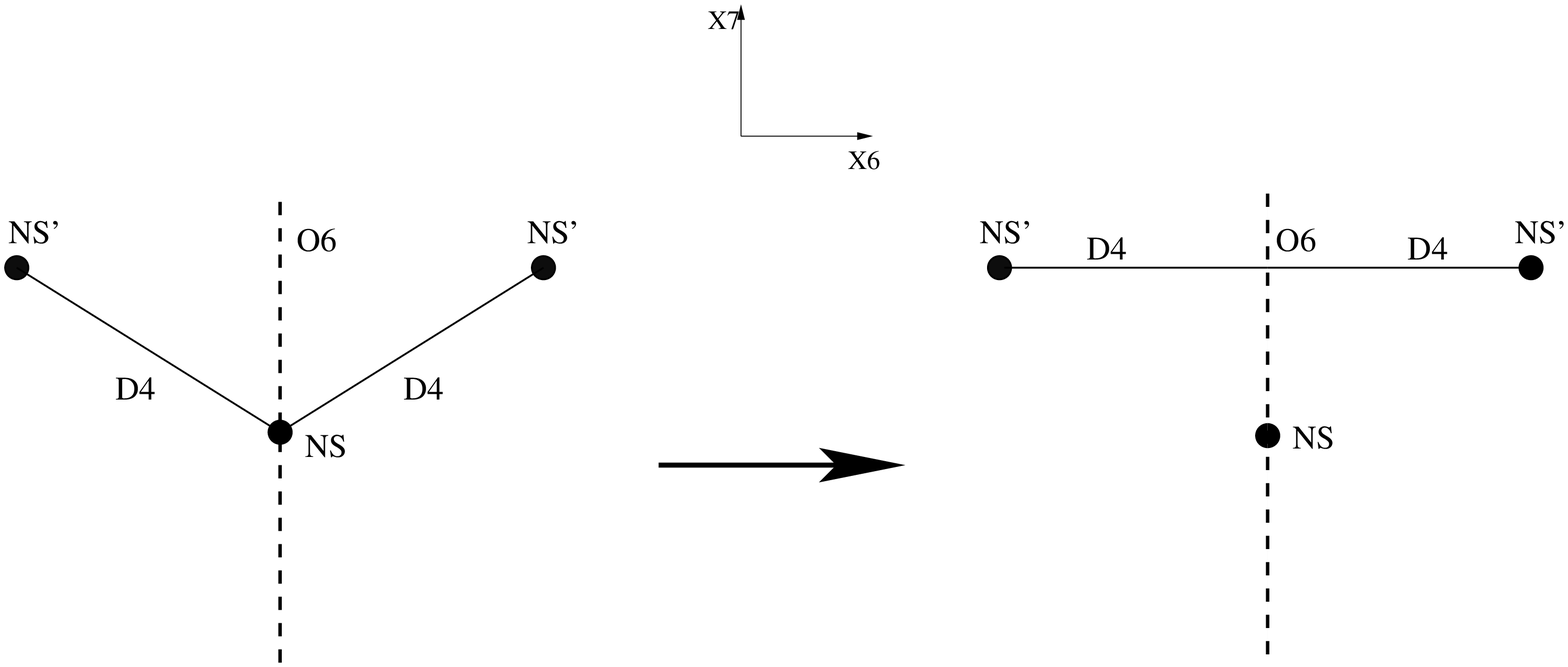}}
\centerline{Fig.\ 5}
\vskip .5cm
The latter configuration is definitely more stable than the
original one, since it preserves space-time supersymmetry.
Combining together field theory and geometry we conclude that when
the $D4$ branes meet the orientifold at positive $x^7$, the
resulting low energy gauge group on them is $SO(N)$.  Now let
$\alpha$ be negative. According to \mn\ the scalar component of $
Q$ becomes tachyonic and gets non zero expectation value. For even
$k$, we have a similar phenomenon. $ Q$ is also a symmetric field
and the resulting gauge symmetry is again $SO(N)$. Geometrically
this $SO(N)$ occurs when the $D$ branes detach from the $NS$ pile
meeting the orientifold at $x^7<0$ (see fig.6).

\vskip 1cm \centerline{\epsfxsize=100mm\epsfbox{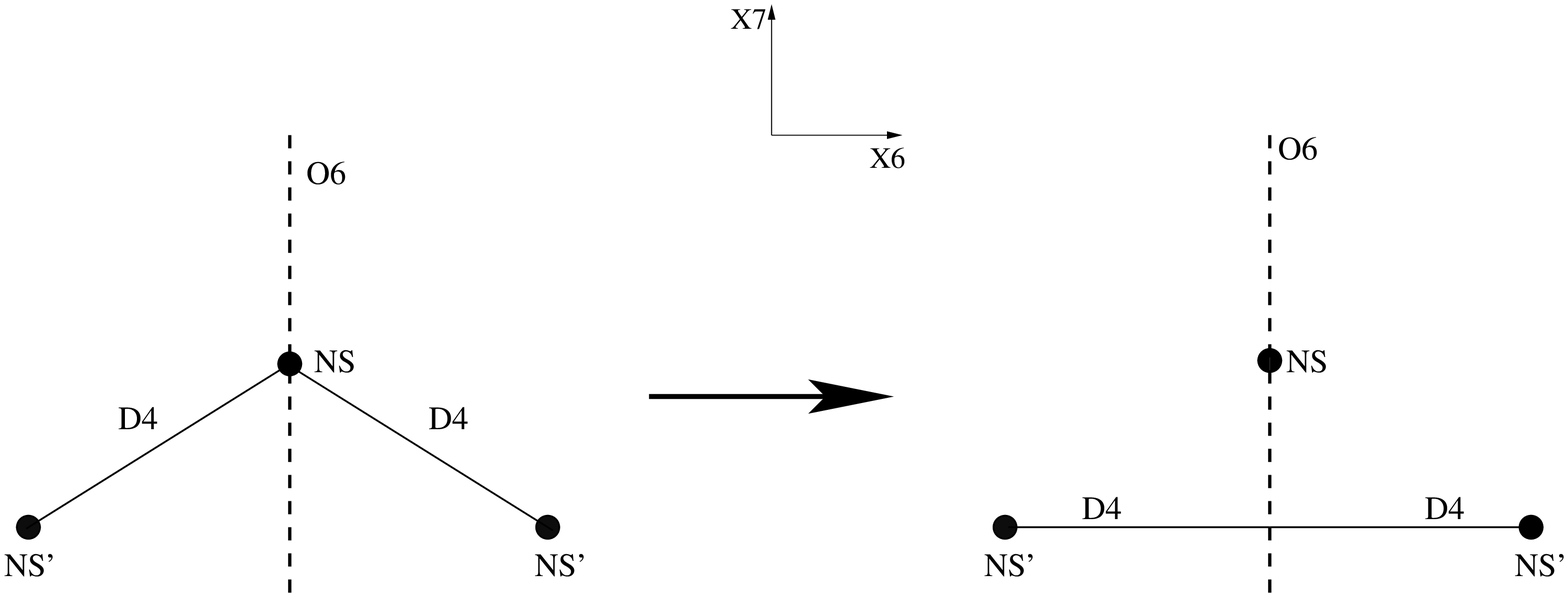}}
\centerline{Fig.\ 6}
\vskip .5cm
For generic values of $\alpha$ the theory on the $D4$ branes is
$SO(N)$, at the critical point $\alpha=0$ there is  an enhancement
of the gauge symmetry to $U(N)$.

If $k$ is odd the picture is different. Unlike $\tilde Q$, the
field $Q$ is antisymmetric. Its expectation value breaks the
$U(N)$ group down to $Sp({N\over 2})$. Now the enhanced symmetry
point $\alpha=0$ is a phase transition from an $SO(N)$ phase for
positive $\alpha$ to a different, $Sp({N\over 2})$ phase for
negative $\alpha$. Geometrically it means that $D4$ branes meeting
the orientifold at $x^7>0$ have $SO(N)$ gauge group at low energy,
while those that meet the orientifold at $x^7<0$ have $Sp({N\over
2} )$ as their gauge group.

\lref\bh{ J.~H.~Brodie and A.~Hanany,
Nucl.\ Phys.\ B {\bf 506}, 157 (1997) [arXiv:hep-th/9704043].
}

We have then another derivation of the phenomenon of the change of
sign of an orientifold when it passes through an odd number,
larger than $1$, of $NS$ branes. This is in full accordance with
the results of \lll\ and \tsabar\ for the case $k=1$.

As explained in these papers this background with an odd $k$ is
not stable by itself. From string theory point of view it has a
non zero tadpole for $RR$ form. From field theory perspective the
matter content of a symmetric and a conjugate antisymmetric chiral
fields in $U(N)$ theory is anomalous. It can be stabilized though
by adding $8$ half $D6$ branes parallel to the $O6$ along half of
the $x^7$ axis, ending on the $NS$ branes. According to \bh\ the
$4-6$ open strings between the $D4$ branes and those half $D6$
branes provide the chiral matter required for cancelling the
anomaly. We do not deal with these phenomena here.

 A parallel
argumentation leading to the same conclusions uses \prad,
\brunner, \bach\ and \huss. Like in sec. 2.3, we can think of the
branes as being continuously rotated by angle
$\beta={\pi-\alpha\over 2}$ from the positive $x^7$ axis (fig. 7).

\vskip 1cm \centerline{\epsfxsize=35mm\epsfbox{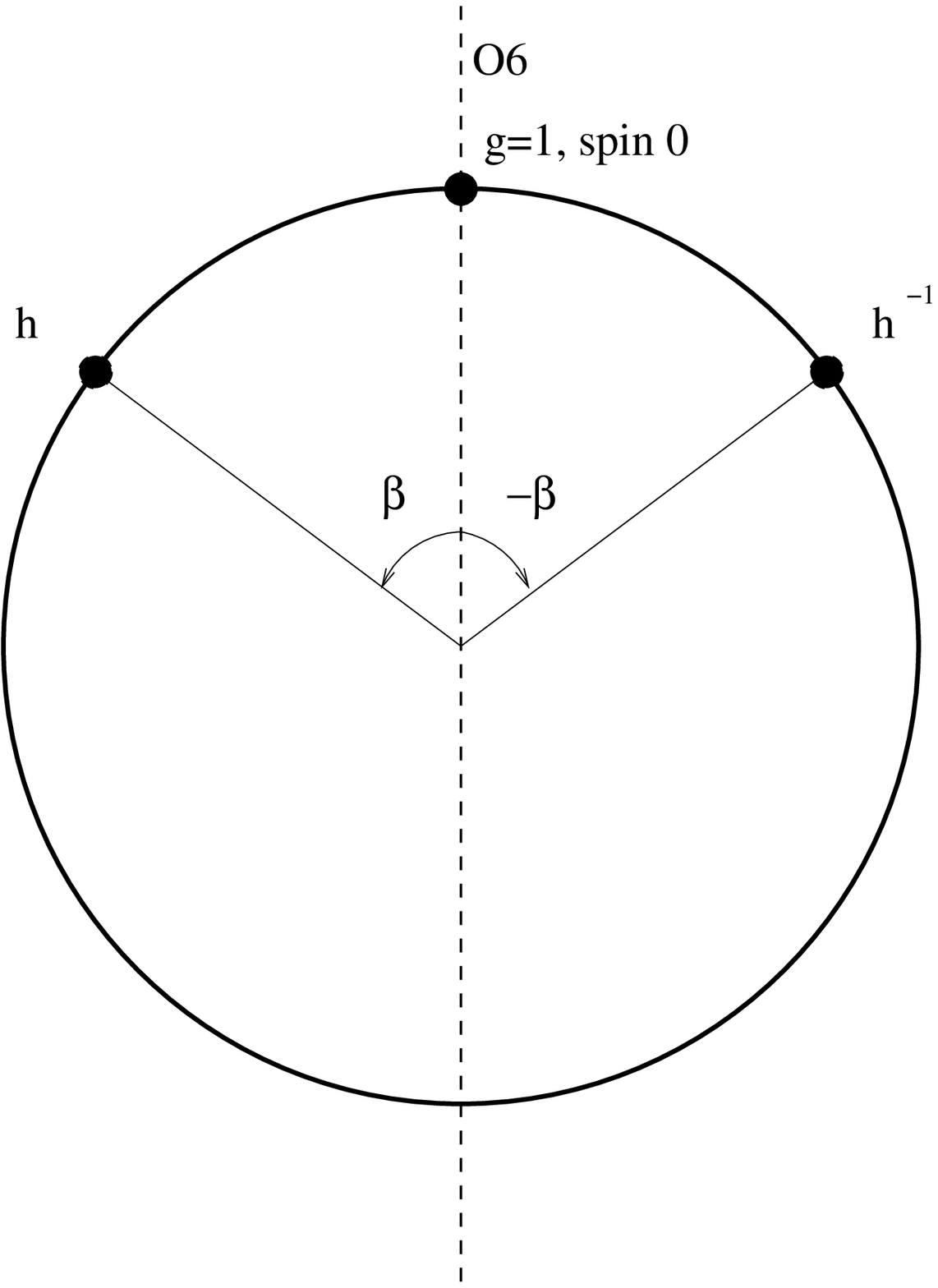}}
\centerline{Fig.\ 7}
\vskip .5cm

 For $\beta=0$ the two
stacks of $D$-branes coincide at $g=1$. They correspond to the
primary field of spin $0$, so the open strings stretched between
them belong to the same representation. The $SU(2)$ part of the
vertex operator corresponding to the field $Q$,
$V^{B(i,j)}_{-{k-2\over 2}}V^F_{-1}$, has for $\beta=0$, zero
$J^3$ charge and zero dimension, hence it is to be identified with
the primary field of this representation. Since the branes
correspond to an integer spin representation, by \prad, \brunner,
\bach\ and \huss, it is preserved by the orientifold projection
giving rise to a symmetric Chan-Paton configuration. Instead, the
$D$-branes can be thought of as rotated  by an angle
$\gamma={\pi+\alpha\over 2}$ from the negative $x^7$ axis (fig.
8).

\vskip 1cm \centerline{\epsfxsize=35mm\epsfbox{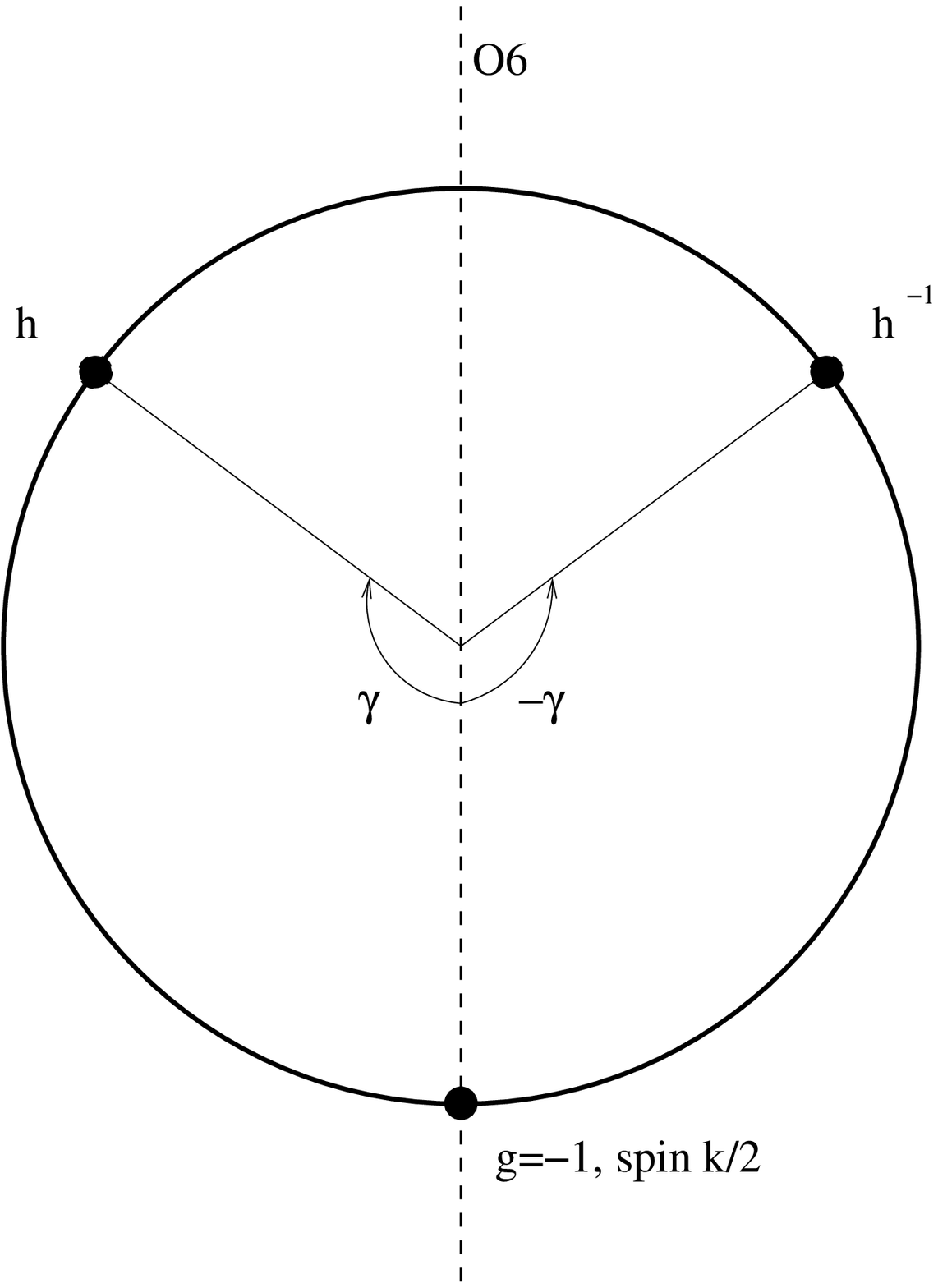}}
 \centerline{Fig.\ 8}
\vskip .5cm

 For $\gamma=0$ the two stacks
coincide at $g=-1$, corresponding to the spin ${k\over 2}$
representation. The open strings between them are again in the
spin $0$ representation. For $\gamma=0$ the $SU(2)$ part of the
vertex operator corresponding to the field $Q$,
$V^{B(i,j)}_{{(k-2)\over 2}}V^F_{1}$, is the one with zero charge
and mass. This is then the primary field corresponding to this
representation. If $k$ is odd, the spin ${k\over 2}$ is half
integral, \prad, \brunner, \bach, \huss\ then imply that the
orientifold projection takes $Q$ to minus itself, giving rise to a
Chan-Paton antisymmetric field. For $0\le\beta<{\pi \over 2}$ the
field $\tilde Q$ is tachyonic, by \mp, breaking the symmetry to
$SO(N)$. The field $Q$ is then massive. For ${\pi\over 2}
<\beta\le\pi$ the field $ Q$ is tachyonic. If $k$ is odd, the
breaking then is to $Sp({N\over 2})$.

\newsec{Conclusions}

The regularized background of $k$ parallel $NS$ branes was used to
study the effect of such branes on the $RR$ charge of an
orientifold. The main tool is the alternating sign of orientifold
projection among open strings members of  $SU(2)$ multiplet found
in \vev\ and \vod. The careful study of the current modes
preserved by the $D$ branes and of their orientifold
transformation, leading to eq. \orgtt\ and \orgtpm, is needed to
establish this sign alternation. It was also shown to be
consistent with the signs assigned in  \prad, \brunner,\bach\  and
\huss\ to Mobius diagrams on $SU(2)$ branes. In principle one may
also obtain this results expressing the vertex operator in terms
of the scalar field $Y$ defined in eq. \y\ as in \gor\ and
studying the orientifold behavior of this field  . This analysis
is deferred for a possible future publication. In \gor,
instabilities were identified in a system of $2$ stacks of $D4$
branes ending on a pile of $k$ $NS$ branes for a generic angle
between them. Introducing an orientifold into such a system and
using the results of sec. 2, these instabilities could be shown to
break the gauge symmetry on the $D$ branes either to an orthogonal
or to a symplectic group. The choice between these two
possibilities characterizes the sign of the orientifold. A change
in the sign was identified for odd $k$. This is a generalization
of the analysis of \lll\ and \tsabar\ for $k>1$. In this way we
could follow the change in the nature of the orientifold as a
phase transition in the gauge theory on the $D$ branes which end
on the pile of $NS$ branes.

\vfill\eject

\bigskip
\noindent{\bf Acknowledgements:} We thank E. Rabinovici for
suggesting the subject and for discussions. We also thank M.
Berkooz and A. Giveon for discussions. This work is supported in
part by the Israel Academy of Sciences and Humanities - Centers of
Excellence program, The German-Israel Bi-national Science
Foundation and the European RTN network HPRN-CT-2000-00122. S.E.
thanks the TH division at CERN for hospitality while part of this
work was done. \listrefs
\end